\documentclass[a4paper,
                keeplastbox,   
              ]{jacow}
\makeatletter%
	\ifboolexpr{bool{xetex}}
	 {\renewcommand{\Gin@extensions}{.pdf,%
	                    .png,.jpg,.bmp,.pict,.tif,.psd,.mac,.sga,.tga,.gif,%
	                    .eps,.ps,%
	                    }}{}
\makeatother

\ifboolexpr{bool{xetex} or bool{luatex}} 
 {}                                      
 {\usepackage[utf8]{inputenc}}           

\usepackage[USenglish]{babel}			 

\usepackage[final]{pdfpages}
\usepackage{multirow}
\usepackage{ragged2e}

\ifboolexpr{bool{jacowbiblatex}}%
 {%
  \addbibresource{jacow-test.bib}
  \addbibresource{biblatex-examples.bib}
 }{}
\begin{document}

\title{Numerical Simulations to Evaluate and Compare the Performances of Existing and Novel Degrader Materials for Proton Therapy}

\author{R.~Tesse\thanks{robin.tesse@ulb.ac.be}, A.~Dubus, N. Pauly \\ 
		Service de Métrologie Nucléaire, Université Libre de Bruxelles, Brussels, Belgium \\
		C.~Hernalsteens, W.~Kleeven, F.~Stichelbaut \\
		Ion Beam Applications (IBA), Louvain-la-Neuve, Belgium}
 
\maketitle

\begin{abstract}
The performance of the energy degrader in terms of beam properties directly impacts the design and cost of cyclotron-based proton therapy centers. The aim of this study is to evaluate the performances of different existing and novel degrader materials. The quantitative estimate is based on detailed GEANT4 simulations that analyze the beam-matter interaction and provide a determination of the beam emittance increase and transmission. Comparisons between existing (aluminum, graphite, beryllium) and novel (boron carbide and diamond) degrader materials are provided and evaluated against semi-analytical models of multiple Coulomb scattering. The results showing a potential in emittance reduction for novel materials are presented and discussed in detail.
\end{abstract}

\section{Introduction}
Cyclotron-based proton therapy installation often rely on a degrader system to modulate the beam energy from the cyclotron output energy to the energy necessary during the clinical treatment. The degrader also induces a growth in the transverse phase space and an increase of the energy spread. Both effects will negatively impact the transmission in the extraction beamline and through the gantry. This increases the beam losses, leading to shielding and building activation, and impacts the beam properties at isocenter. The transverse properties of the beam are indeed crucial for a detailed understanding of installations in operation and for the design of future systems, in particular with Pencil Beam Scanning treatment modes where a small spot size at isocenter is favored \cite{item:pbs}. Similarly, a precise knowledge of the beam loss pattern is of importance for the design of the shielding surrounding the installation, in particular for compact systems. A complete model of the \emph{Proteus~One} system using a realistic geometry of the extraction line, degrader and gantry and a complete description of the beam transport is being developed, which uses GEANT4 for a Monte-Carlo (MC) evaluation of the beam-matter interactions. The accuracy in the evaluation of the degrader effects is of primary importance. It can be assessed by comparing different MC models and analytical models (see \cite{item:nimb} for a comparison including GEANT4 and other MC codes made in the framework of shielding studies). 

\begin{figure}[!ht]
    \centering
    \includegraphics*[width=0.2\textwidth]{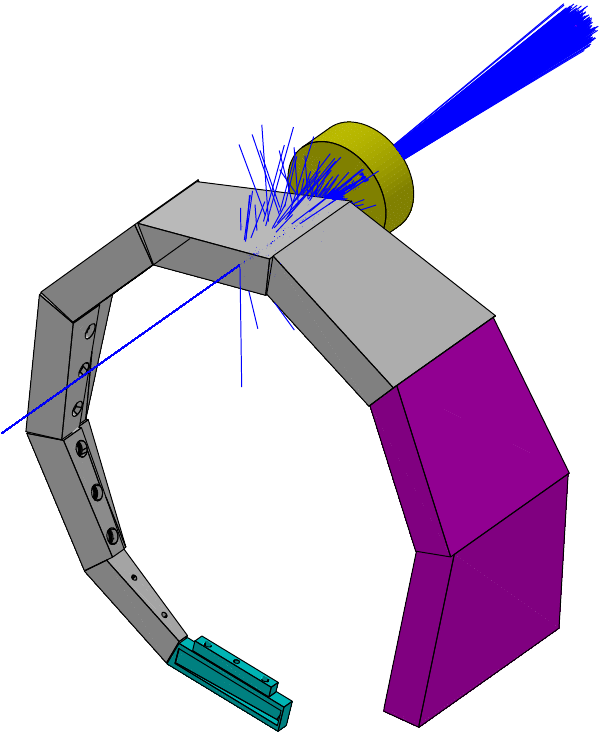}
    \caption{Realistic model of the IBA energy degrader used for GEANT4 simulations combining three degrader materials (aluminum (green), graphite (gray) and beryllium (purple)). The downstream circular collimator is also shown (yellow). A sample of proton tracks are shown in blue.}
    \label{fig:wheel}
\end{figure}

In this work we apply those tools and methods to evaluate and compare the performances of different degrader materials. The materials that we consider are those of the currently used degrader of the IBA systems as well as boron carbide and diamond. The IBA degrader is made of wedge-shaped slabs of materials wrapped around a rotating wheel (see Fig.~\ref{fig:wheel}). The realistic geometry, as used in our GEANT4 simulations, is shown in Fig.~\ref{fig:wheel}. It makes use of aluminum, and beryllium for degradation to the highest, medium and lowest energies, respectively. Boron Carbide has been considered as a material that could reduce the emittance growth and increase the transmission efficiency (see Ref.\,\cite{item:1}). Diamond, which has recently been proposed as a novel degrader material to reduce the emittance growth by one of the author (W.K.), is also evaluated. The densities of the different materials are summarized in Table~\ref{table:1}.

\begin{table}[!hbt]
   \centering
   \caption{Density of the Evaluated Materials}
   \label{table:1}
   \begin{tabular}{lc}
       \toprule
       \textbf{Material}      & \textbf{Density}  \\
       \midrule
           Aluminum         & \SI{2.7}{g/cm^3}   \\
           Graphite         & \SI{1.7}{g/cm^3}   \\
           Beryllium        & \SI{1.9}{g/cm^3}   \\
           Diamond          & \SI{3.5}{g/cm^3}    \\
           Boron carbide    & \SI{2.5}{g/cm^3}    \\
       \bottomrule
   \end{tabular}
   \label{l2ea4-t1}
\end{table}

\section{Tools and methods}
The beam interactions in the degrader material are of electromagnetic (energy loss and Coulomb scattering) and nuclear origins. A major effect of the nuclear interactions (elastic and inelastic) is to create heavier tails in the phase space projected distributions than the Multiple Coulomb scattering (MCS). This introduces complications if one aims at comparing detailed MC results with semi-analytical MCS models. In order to precisely interpret the results obtained with GEANT4 and compare them with the Fermi-Eyges transport theory \cite{item:Eyges}), we propose and discuss a method to find the cuts that will optimally restore a Gaussian distribution, in turn allowing to rigorously compare the distributions using rms quantities.

The Fermi-Eyges transport theory\cite{item:Eyges} provides the covariance matrix of the phase-space distribution at any given depth $z$ in the material. Supposing a pencil beam at the entrance of the material, the probability density in $x$ (transverse displacement) and $\theta$ (transverse angle) is given by
\begin{equation}\label{eq:Eyges_proba}
P(x,\theta) dx d\theta = \frac{1}{2\pi\sqrt{A_0A_2-A_1^2}}e^{-\frac{1}{2}\frac{A_0x^2-2A_1x\theta+A_2\theta^2}{A_0A_2-A_1^2}} dxd\theta\,.
\end{equation}

The quantities $A_i$ are the $i^{th}$ moments of the scattering power $T$\cite{item:gottschalk}.
\begin{align}
A_i(z) &= \int_0^{z}(z-u)^iT(u) du\,.
\end{align}

The $\theta$ (resp. $x$) distribution is obtained by integration of Eq.\eqref{eq:Eyges_proba} over $x$ (resp. $\theta$). With this formalism, the beam properties at the exit of any material can be deduced from the scattering power alone ($\left\langle \theta^2 \right\rangle = A_0$, $\left\langle x \theta\right\rangle = A_1$, $\left\langle x^2\right\rangle = A_2$ and the emittance $\epsilon = \pi \sqrt{A_0A_2-A_1^2}$). Different forms for the scattering power have been proposed \cite{item:farley, item:gottschalk}. We use the one by Gottschalk \cite{item:gottschalk}.

\begin{figure}[!ht]
    \centering
    \includegraphics*[width=0.45\textwidth]{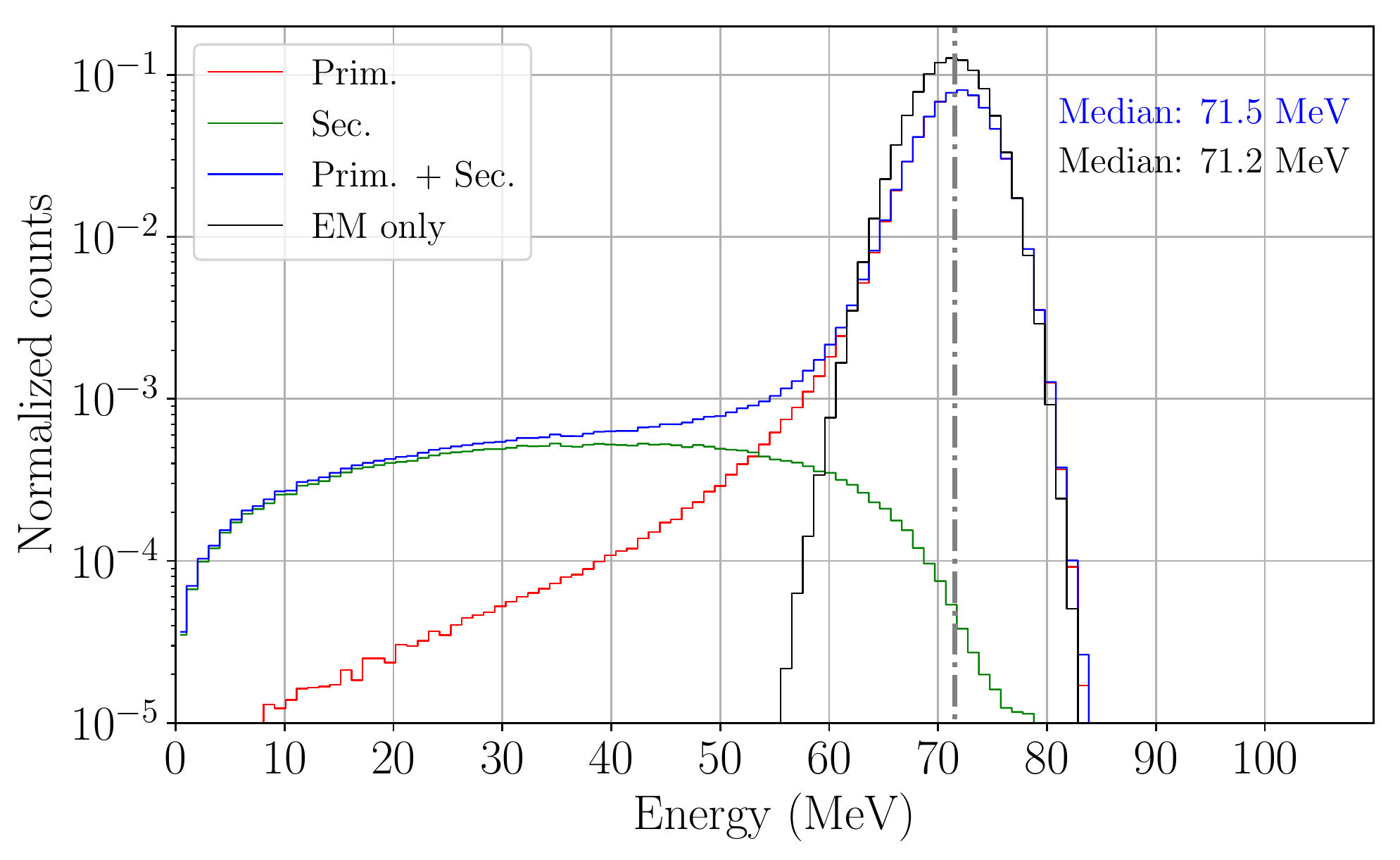}
    \caption{Energy spectrum for beryllium degrading from 230 MeV to 70 MeV.}
    \label{fig:energy_spectrum_secondaries_Beryllium}
\end{figure}

\begin{figure}[!ht]
    \centering
    \includegraphics*[width=0.45\textwidth]{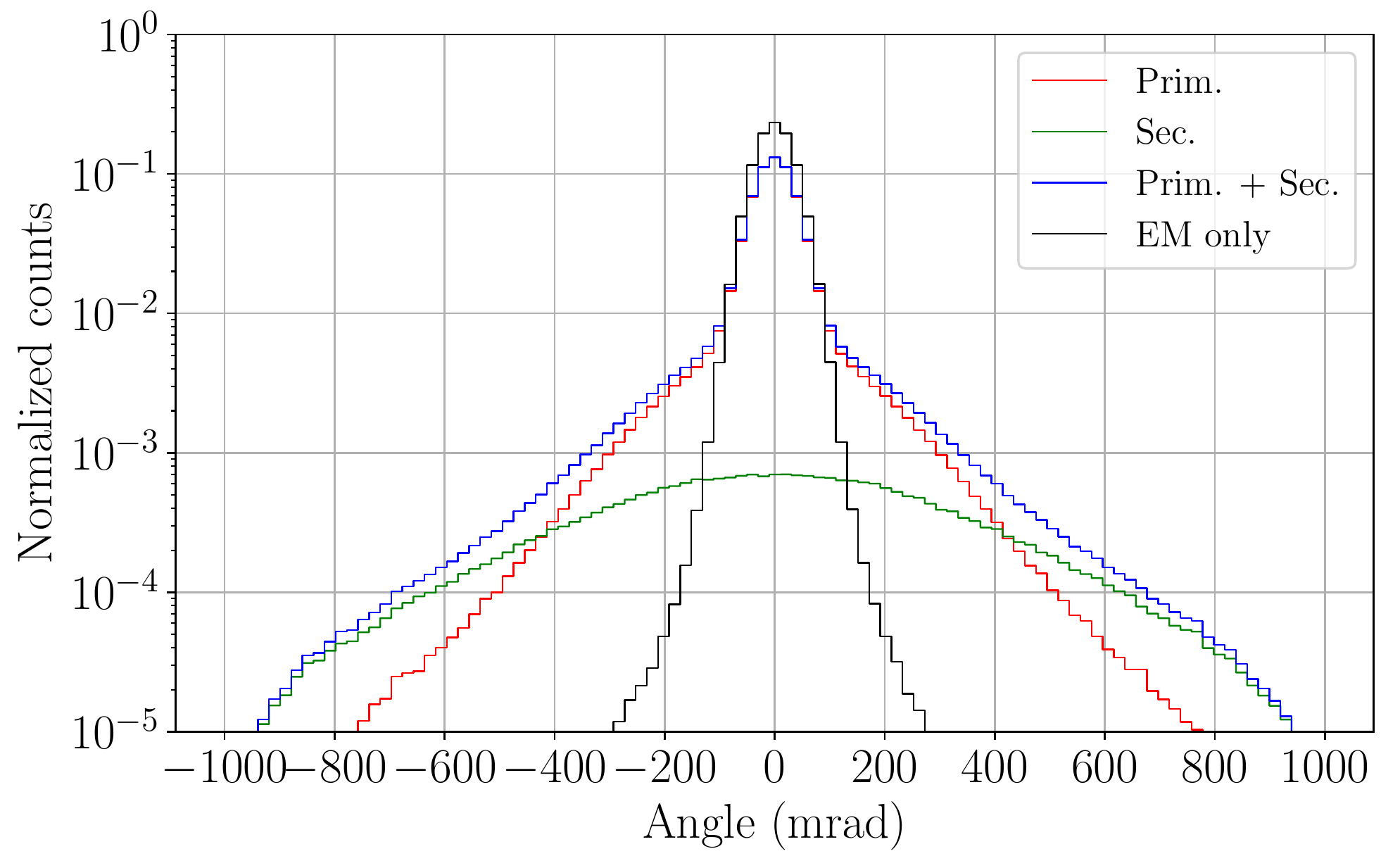}
    \caption{Angular distribution for beryllium degrading from 230 MeV to 70 MeV.}
    \label{fig:angular_spectrum_secondaries_Beryllium}
\end{figure}

The Monte Carlo simulations are performed with GEANT4 (v.10.03) using $10^6$ primaries. To perform the comparisons between materials, a simple geometry which consists of a semi-infinite block is first used. The post processing distinguishes between primary particles and secondaries produced by inelastic nuclear interactions. As shown in Fig.~\ref{fig:energy_spectrum_secondaries_Beryllium} and \ref{fig:angular_spectrum_secondaries_Beryllium}, more than two orders of magnitude separate the primaries and secondaries in the peak region (median value) in the energy spectrum. The present study focuses on the transverse distributions and discards those secondaries. 

\begin{figure}[!ht]
    \centering
    \includegraphics*[width=0.45\textwidth]{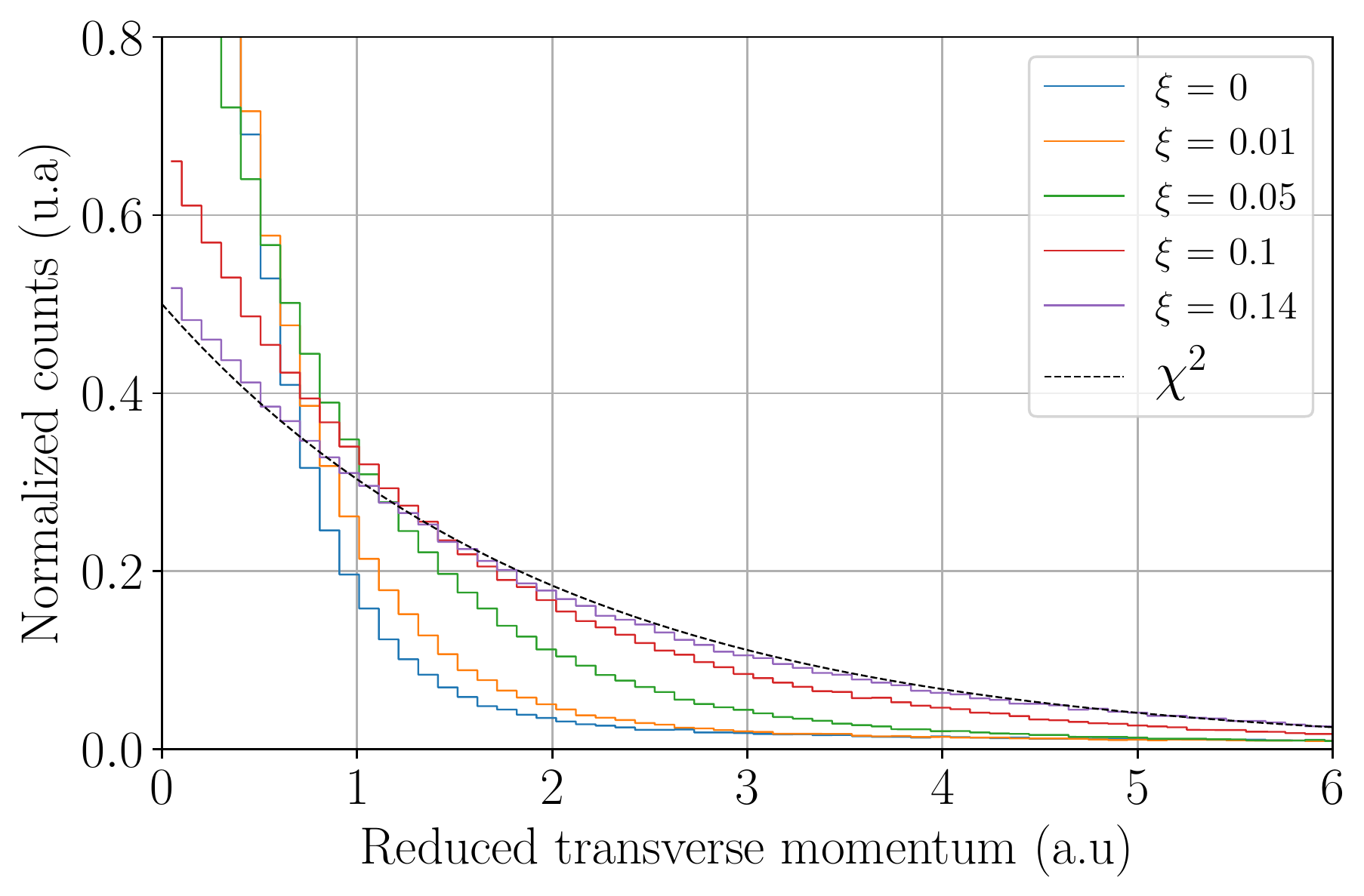}    
    \caption{Reduced $\chi^2$ distribution of the transverse momentum $\tilde{P}_T$ for different cut values (see text), for a beryllium block degrading from 230 to 70 MeV.}
    \label{fig:QuantileChi_distribution}
\end{figure}

In order to compare the GEANT4 simulations with the Fermi-Eyges estimates we must be able to separate the nuclear halo contribution from the core. The method we use optimally cuts the distribution to recover a Gaussian core without introducing arbitrary biases. The transverse momentum $\tilde{P}_T$ of a Gaussian core constructed from the reduced angular distributions ($\tilde{\theta}_{x,y} = \theta_{x,y}/\sqrt{\langle (\theta_{x,y}-\langle \theta_{x,y} \rangle)^2 \rangle}$) follows a $\chi^2$ distribution. We apply cuts on the $P_T$ distribution until $\sigma_{\tilde{P}_T} = 2$ and therefore recover projected distributions which are Gaussian. By construction this also ensures that no difference in the treatment of the two transverse planes is introduced, also a key point when comparing with analytical models. This is illustrated in Fig.~\ref{fig:QuantileChi_distribution} for a beryllium block degrading the beam from 230 to 70 MeV. In that particular case the optimum Gaussian cut (denoted $\xi$) that removes the nuclear halo is $\xi = 0.142$. Nuclear scattering is thus a non-negligible contributor, however, our method allows to disentangle the halo and the core. The effect of the cut is accounted for by adding $\xi$ in the computation of the losses.

The Fermi-Eyges angular distribution approximates the Molière treatment of the MCS. In order to directly compare GEANT4 distributions with Fermi-Eyges estimates we also ran the simulations using a reduced physics list including only EM interactions. Figures~\ref{fig:energy_spectrum_secondaries_Beryllium} and \ref{fig:angular_spectrum_secondaries_Beryllium} show those distributions (black). We apply the same method in that case, however the single scattering tails of the EM distribution are much weaker than the tails of the nuclear halo ($\xi_{EM} = 0.85 \%$).

\section{Comparison of materials}

\begin{figure}[!ht]
    \centering
    \includegraphics*[width=0.45\textwidth]{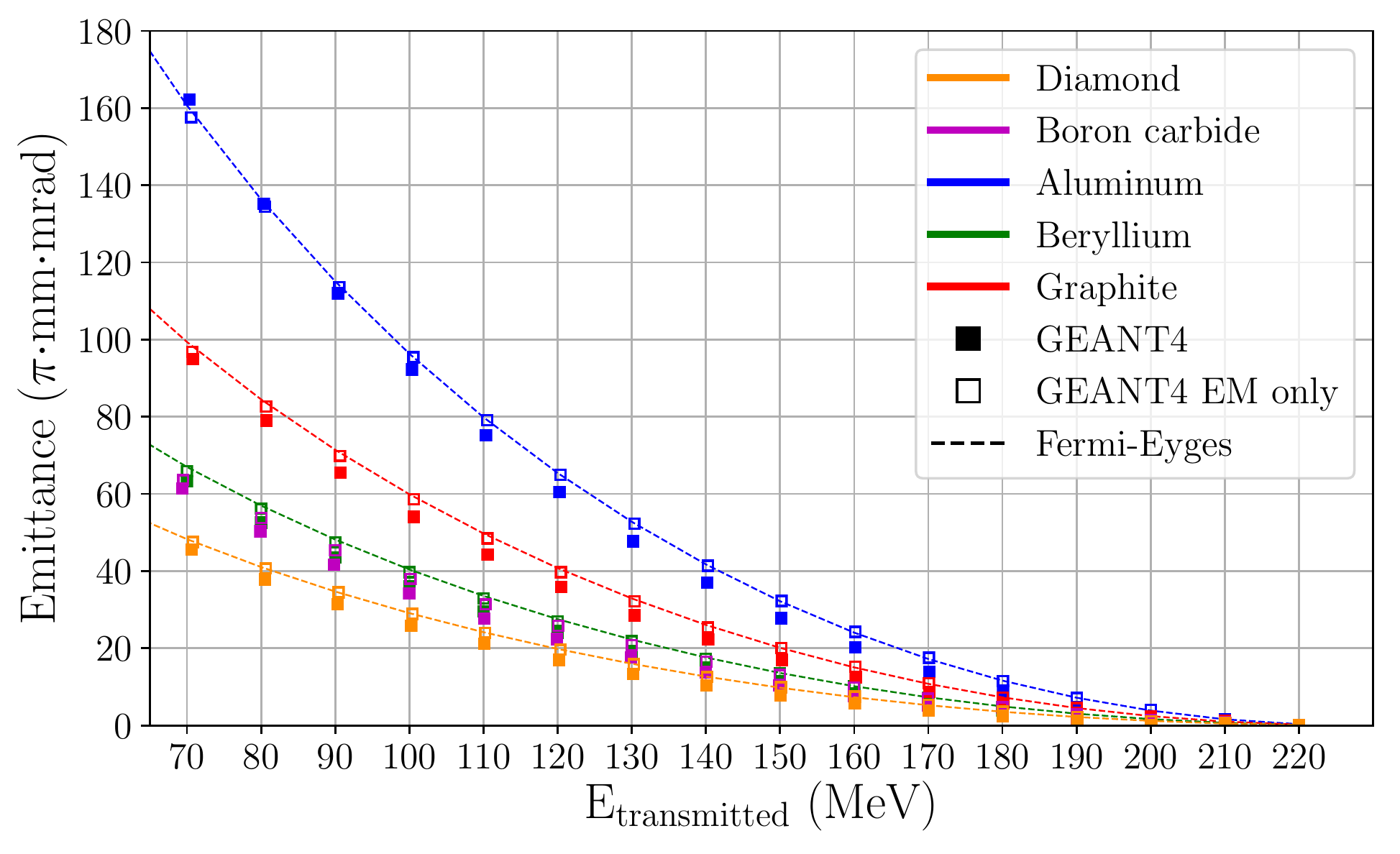}    
    \caption{Comparison of the emittance values obtained using the FE model (except for boron carbide), GEANT4 results for the core only (cuts applied) and GEANT4 results obtained with a EM-only physics list. The agreement between the FE model and the GEANT4 (EM) simulations is clearly visible.}
    \label{fig:emittance_materials}
\end{figure}

The emittances are shown in Fig.~\ref{fig:emittance_materials}. The values obtained from the Fermi-Eyges model (continuous curves) are in excellent agreement with the GEANT4 results obtained with a physics list containing only the electromagnetic interactions (\textsc{G4EmPenelopePhysics}, empty markers). The results from the GEANT4 simulations using the complete physics list (\textsc{qgsp\_bic}) and processed following the method described above are observed to be lower than the EM models in all cases. We interpret this by the fact that the elastic nuclear scattering are favored at large angles, resulting in an angular distribution which has a wider non-Gaussian shape. By recovering only the Gaussian core of this distribution, the final rms estimates are indeed lower. These MC simulations can serve as inputs for constructing rms beam distributions (beam $\Sigma$ matrix) to be used in beam transport codes. In that case, those results show that it is essential to consider a model including the nuclear interactions because they lead to a correct representation of the beam losses arising from inelastic scattering (losses removed by filtering out the secondaries) but also from nuclear elastic scattering at large angles (losses removed by applying optimal Gaussian cuts). By using the rms values following that method, the Gaussian core is correctly represented but heavier tails are not taken into account. However, using rms estimates from the purely EM distributions would overestimate the width of the core distribution.

\begin{figure}[!ht]
    \centering
    \includegraphics*[width=0.5\textwidth]{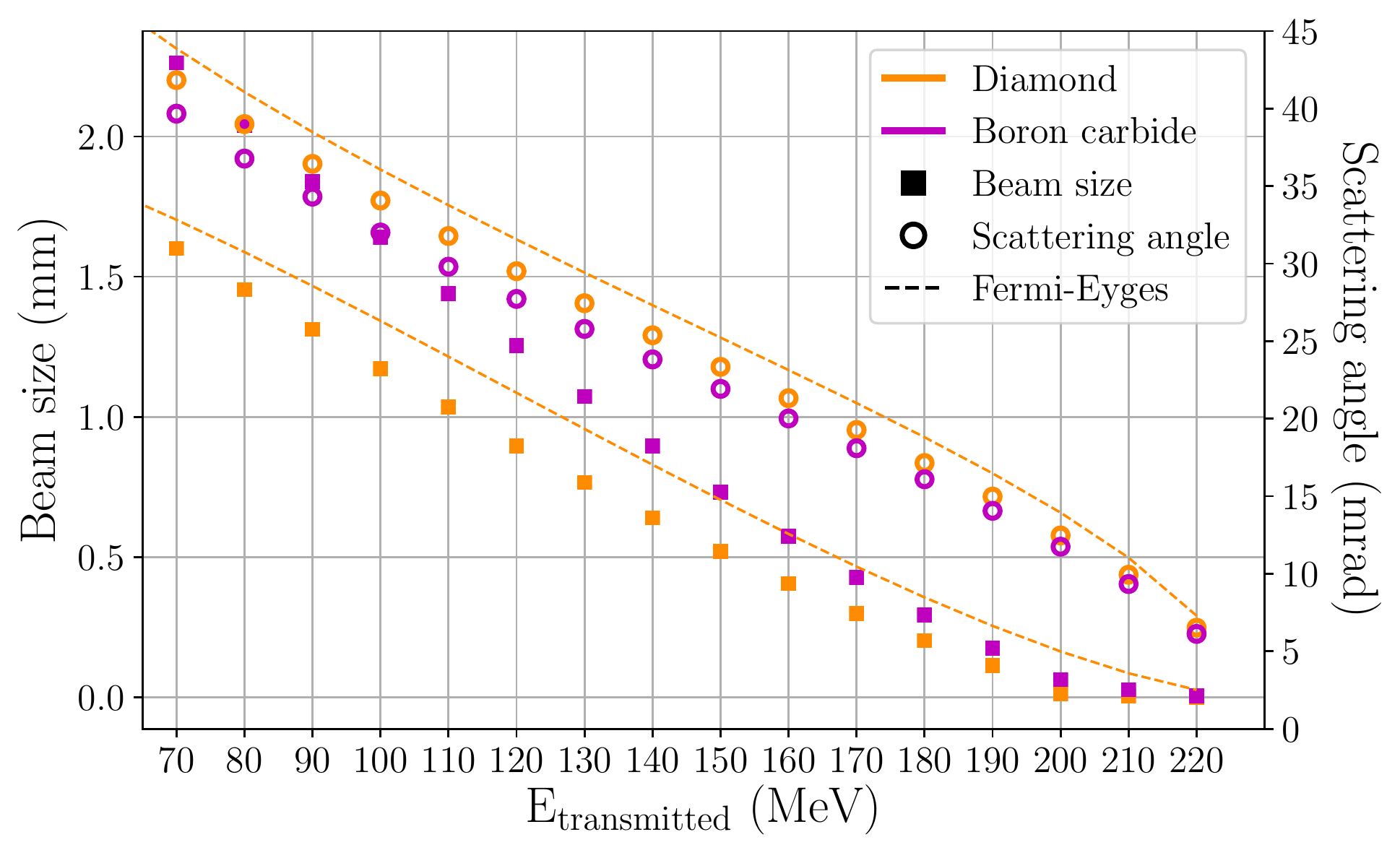}    
    \caption{Beam size (squares) and scattering angle (circles) for diamond and boron carbide explaining the behavior of the emittance (Fig.~\ref{fig:emittance_materials}). The GEANT4 simulations (with cuts) including nuclear effects are shown (markers) together with the FE model (dashed curves) for diamond.}
    \label{fig:emittance_materials2}
    \vspace*{-\baselineskip}
\end{figure}

For all models, the differences in emittance growth are clearly observed. In particular, the behaviors of aluminum, graphite and beryllium follow the expected order: the emittance increase is minimized by a high material density and by a low atomic number $Z$. This explains why beryllium is a material of choice and why it is used in the IBA degrader at low energy. The results obtained for a diamond degrader are particularly interesting: a 30\% reduction is observed for the lowest energies. Figure~\ref{fig:emittance_materials2} shows the beam size ($\langle x^2 \rangle$) and angular spread ($\langle \theta^2 \rangle$) for diamond and boron carbide. The reduced emittance obtained with diamond arise from the high material density, thus reducing the required slab length and reducing the beam size increase as the beam propagates in the slab. This effect dominates the large scattering angle of carbon compared to beryllium leading to a larger angular spread ($\langle \theta^2 \rangle$) than beryllium and boron carbide. Figure~\ref{fig:emittance_materials2} also illustrates the trade-off between beam size and angular spread characterizing boron carbide: its density ($\rho=2.5\text{g/cm}^3$) is lower than that of diamond ($\rho=3.5\text{g/cm}^3$) leading to a large beam size, however the angular spread is lower.

\section{Conclusion}
The performances of different materials for their use as energy degrader have been evaluated with GEANT4 and a method is proposed to optimally account for the rms properties of the distributions core. It is shown that the elastic nuclear interactions, although introducing scattering at large angles, make the Gaussian core narrower. Our method quantifies the inelastic nuclear losses and the halo effect from nuclear scattering. The results are consistent with purely EM models and confirmed with MC simulations considering only EM interactions. Diamond, as a novel degrader material is evaluated in detail: the results confirm the improved performance of a 30\% decrease in emittance at low energies compared with beryllium which is confirmed by the relative impact of diamond on beam size and angular spread.

\section{Acknowledgement}
Work by R. Tesse was supported by a research grant from Association Vinçotte Nuclear (AVN).

\end{document}